\documentclass{article}
\usepackage{cite}
\usepackage{graphicx}
\usepackage{dcolumn}

\begin{document}

\date{}
\title{Comment on: ``Some quantum aspects of a particle with electric quadrupole
moment interacting with an electric field subject to confining potentials''.
Int. J. Mod. Phys. A 29 (2014) 1450117}
\author{Francisco M. Fern\'{a}ndez\thanks{%
fernande@quimica.unlp.edu.ar} \\
%EndAName
INIFTA, DQT, Sucursal 4, C.C 16, \\
1900 La Plata, Argentina}
\maketitle

\begin{abstract}
We analyze the results obtained from a model consisting of the interaction
between the electric quadrupole moment of a moving particle and an electric
field. We argue that the system does not support bound states because the
motion along the $z$ axis is unbounded. It is shown that the author obtains
a wrong bound-state spectrum for the motion in the $x-y$ plane and that the
existence of allowed cyclotron frequencies is an artifact of the approach.
\end{abstract}

In a paper published in this journal Bakke\cite{B14b} studies the bound
states of a quantum-mechanical model given by the interaction between the
electric quadrupole of a moving particle and an electric field. In two other
models the author adds a harmonic potential and a linear plus harmonic
potential. By means of suitable transformations of the time-dependent
Schr\"{o}dinger equation the author derives an eigenvalue equation for the
radial part of the wavefunction that can be treated by means of the
Frobenius (or power-series) method. In the two cases with radial potentials
the power-series method leads to three-term recurrence relations. Suitable
truncation of the series yields analytical expressions for the energy
eigenvalues and a striking result: the appearance of allowed cyclotron
frequencies. The purpose of this Comment is the analysis of the truncation
method used by Bakke and its effect on the physical conclusions drawn in his
paper.

Upon taking into account the interaction between the electric quadrupole of
the moving particle and the electric field the author derives the
time-dependent Schr\"{o}dinger equation
\begin{equation}
i\frac{\partial \psi }{\partial t}=-\frac{1}{2m}\left( \frac{\partial ^{2}}{%
\partial \rho ^{2}}+\frac{1}{\rho }\frac{\partial }{\partial \rho }+\frac{1}{%
\rho ^{2}}\frac{\partial ^{2}}{\partial \varphi ^{2}}+\frac{\partial ^{2}}{%
\partial z^{2}}\right) \psi -\frac{QE_{0}}{\rho }\psi ,
\label{eq:Schro_fime_dep}
\end{equation}
in cylindrical coordinates ($0\leq \rho <\infty $, $0\leq \varphi \leq 2\pi $%
, $-\infty <z<\infty $) and units such that $\hbar =c=1$ (we have recently
criticized this kind of nonrigorous choice of suitable units\cite{F20}). The
meaning of every parameter in this equation is given in the author's paper%
\cite{B14b}.

Since the Hamiltonian operator is time-independent and commutes with $\hat{p}%
_{z}$ and $\hat{L}_{z}$ the author looks for a particular solution of the
form
\begin{equation}
\psi (t,\rho ,\varphi ,z)=e^{-i\mathcal{E}t}e^{il\varphi }e^{ikz}R(\rho ),
\label{eq:psi}
\end{equation}
where $l=0,\pm 1,\pm 2,\ldots $ and $-\infty <k<\infty $. The function $%
R(\rho )$ satisfies the differential equation
\begin{equation}
R^{\prime \prime }+\frac{1}{\rho }R^{\prime }-\frac{l^{2}}{\rho ^{2}}R+\frac{%
\tilde{Q}E_{0}}{\rho }R+\zeta ^{2}R=0,  \label{eq:dif_eq_R_1}
\end{equation}
where $\zeta ^{2}=2m\mathcal{E}-k^{2}$ and $\tilde{Q}=2mQ$. After analyzing
the behaviour of $R(\rho )$ at infinity the author argues that ``Therefore,
we can find either scattering states $\left( R\cong e^{i\zeta \rho }\right) $
or bound states $R\cong \left( e^{-\tau \rho }\right) $. Our intention is to
obtain bound state solutions, then we consider $\zeta ^{2}=-\tau ^{2}$''.

It is worth mentioning that the Schr\"{o}dinger equation discussed above
does not have bound-states for any value of $\tilde{Q}E_{0}$ since the
motion is unbounded along the $z$ direction (the Hamiltonian operator
commutes with $\hat{p}_{z}$). For this reason one introduces the term $%
e^{ikz}$, and the energy, which depends on $k^{2}/(2m)$, takes all the
values $\mathcal{E}\geq \zeta ^{2}/(2m)$. To be more specific, we have bound
states only if
\begin{equation}
\int \int \int \left| \psi (t,\rho ,\varphi ,z)\right| ^{2}\rho \,d\rho
\,d\varphi \,dz<\infty ,  \label{eq:bound-state_def}
\end{equation}
as shown in any textbook on quantum mechanics\cite{LL65,CDL77}. In all the
examples discussed here the improper integral over $z$ is obviously
divergent. However, since there is physical interest in the bound states for
the motion restricted to the plane $x-y$\cite{LL65,CDL77} we will discuss
them in what follows.

In the second model the author adds the potential $V(\rho )=\frac{1}{2}%
m\omega ^{2}\rho ^{2}$ ($\rho ^{2}=x^{2}+y^{2}$) that is clearly unable to
bound the motion along the $z$ axis but, as stated above, we will focus on
the motion in the $x-y$ plane. By means of the change of variables $\xi =%
\sqrt{m\omega }\rho $ one obtains the radial eigenvalue equation
\begin{equation}
R^{\prime \prime }+\frac{1}{\xi }R^{\prime }-\frac{l^{2}}{\xi ^{2}}R+\frac{%
\alpha }{\xi }R-\xi ^{2}R+\frac{\zeta ^{2}}{m\omega }R=0,
\label{eq:dif_eq_R_2}
\end{equation}
where $\alpha =\frac{\tilde{Q}E_{0}}{\sqrt{m\omega }}$.

On writing the solution $R(\xi )$ as
\begin{equation}
R(\xi )=\xi ^{|l|}e^{-\frac{\xi ^{2}}{2}}\sum_{j=0}^{\infty }a_{j}\xi ^{j},
\label{eq:R_series}
\end{equation}
one obtains the three-term recurrence relation
\begin{eqnarray}
a_{j+2} &=&-\frac{\alpha }{(j+2)(j+1+\theta )}a_{j+1}-\frac{g-2j}{%
(j+2)(j+1+\theta )}a_{j},  \nonumber \\
j &=&-1,0,1,2\ldots ,\;a_{0}=1,\,a_{-1}=0,  \label{eq:rec_rel}
\end{eqnarray}
where $\theta =2|l|+1$ and $g=\frac{\zeta ^{2}}{m\omega }-2-2|l|$.

The author argues as follows: ``Bound state solutions correspond to finite
solutions, therefore, we can obtain bound state solutions by imposing that
the power series expansion (22) or the Heun biconfluent series becomes a
polynomial of degree $n$. Through the expression (23), we can see that the
power series expansion (22) becomes a polynomial of degree $n$ if we impose
the conditions:
\begin{equation}
g=2n\ \mathrm{and}\ a_{n+1}=0,  \label{eq:truncation_cond}
\end{equation}
where $n=1,2,\ldots $.''

It clearly follows from these two conditions that $a_{j}=0$ for all $j>n$;
however, the author's statement is a gross conceptual error because a bound
state simply requires that $R(\xi )$ is square integrable:
\begin{equation}
\int_{0}^{\infty }\left| R(\xi )\right| ^{2}\xi \,d\xi <\infty ,
\label{eq:bound-state_def_xi}
\end{equation}
as shown in any textbook on quantum mechanics\cite{LL65,CDL77}. Therefore,
the truncation condition may not render all the bound states.

For example, when $n=1$ the first condition yields $g=2$ and one obtains a
simple analytical expression for $\mathcal{E}_{1,l}$. The second condition $%
a_{2}=0$ yields $\alpha =\alpha _{1,l}=\pm \sqrt{4|l|+2}$ and the cyclotron
frequency $\omega _{1,l}=\left( \tilde{Q}E_{0}\right) ^{2}/(m\alpha
_{1,l}^{2})$. From the general case (\ref{eq:truncation_cond}) the author
derives an analytic expression for $\mathcal{E}_{n,l}$ corresponding to $%
\omega _{n,l}=\left( \tilde{Q}E_{0}\right) ^{2}/(m\alpha _{n,l}^{2})$.
Accordingly, he argues as follows: ``This means that not all values of the
angular frequency $\omega $ are allowed, but some specific values of $\omega
$ which depend on the quantum numbers $\{n,l\}$; thus, we label $\omega
=\omega _{n,l}$.'' Later on he also states that ``Hence, we have seen in Eq.
(28) that the effects of the Coulomb-like potential induced by the
interaction between an electric field and an electric quadrupole moment on
the spectrum of energy of the harmonic oscillator corresponds to a change of
the energy levels, whose ground state is defined by the quantum number $n=1$
and the angular frequency depends on the quantum numbers $\{n,l\}$. This
dependence of the cyclotron frequency on the quantum numbers $\{n,l\}$ means
that not all values of the cyclotron frequency are allowed, but some
specific values of the harmonic oscillator frequency defined in such a way
that the conditions established in Eq. (25) are satisfied and a polynomial
solution to the function $H(\xi )$ given in Eq. (22) is achieved.''

Both conclusions are wrong as we shall see in what follows. In the first
place, the truncation conditions (\ref{eq:truncation_cond}) only yield some
rather rare bound states, based on polynomial functions, that occur for some
arbitrary particular values of $\alpha $. More precisely, the energy $%
\mathcal{E}_{n,l}$ obtained by the author comes from an eigenvalue of the
differential equation (\ref{eq:dif_eq_R_2}) with $\alpha =\alpha _{n,l}$
while any other energy $\mathcal{E}_{n^{\prime },l^{\prime }}$ comes from an
eigenvalue of that equation with $\alpha =\alpha _{n^{\prime },l^{\prime }}$%
; that is to say: those energies are not members of the same spectrum but
eigenvalues of two different operators. This fact is obvious for anybody
familiar with the exact solutions of any quasi-solvable (or conditionally
solvable) model\cite{D88,CDW00,BCD17} (and references therein).

In order to make the point above clearer we solve the eigenvalue equation (%
\ref{eq:dif_eq_R_2}) for $W=\frac{\zeta ^{2}}{m\omega }$ by means of the
reliable Rayleigh-Ritz variational method that is well known to yield
increasingly accurate upper bounds to all the eigenvalues of the
Schr\"{o}dinger equation\cite{P68} (and references therein). For simplicity
we choose the basis set of (non-orthogonal) functions $\left\{ u_{j}(\xi
)=\xi ^{|l|+j}e^{-\frac{\xi ^{2}}{2}},\;j=0,1,\ldots \right\} $. We verify
the accuracy of these results by means of the powerful Riccati-Pad\'{e}
method\cite{FMT89a}. To begin with, notice that the eigenvalue equation (\ref
{eq:dif_eq_R_2}) exhibits bound states for all real values of $l$ and $%
\alpha $ and not only for the particular values of $\alpha $ considered by
the author. This fact is obvious to anybody familiar with the
Schr\"{o}dinger equation. Therefore, there cannot be any quantization of $%
\alpha $ and, consequently, of the oscillator frequency $\omega $.

In order to facilitate de following discussion we call $W^{(n,l)}=2n+2|l|+2$%
, $n=1,2,\ldots $ the eigenvalues coming from the truncation condition and $%
W_{\nu ,l}(\alpha )$, $\nu =0,1,\ldots $ the actual eigenvalues of equation (%
\ref{eq:dif_eq_R_2}) for given values of $l$ and $\alpha $. It is worth
mentioning that $a_{n+1}\left( g=2n,\alpha \right) =0$ exhibits $n+1$ roots $%
\alpha _{n,l}^{(i)}$, $i=1,2,\ldots ,n+1$. All these roots are real as shown
by a theorem of Child et al\cite{CDW00}. For example, $W^{(1,l)}=4+2|l|$, $%
\alpha _{1,l}^{(1)}=-\sqrt{4|l|+2}$ and $\alpha _{1,l}^{(2)}=\sqrt{4|l|+2}$.
When $\alpha =\alpha _{1,0}^{(1)}=-\sqrt{2}$ the two numerical approaches
mentioned above yield $W_{0,0}=W^{(1,0)}=4$, $W_{1,0}=7.693978891$, $%
W_{2,0}=11.50604238$ for the three lowest eigenvalues. We appreciate that
the truncation method only yields the lowest eigenvalue with $l=0$ and
misses all the other ones. When $l=1$ and $\alpha =\alpha _{1,1}^{(1)}=-%
\sqrt{6}$ both numerical methods yield $W_{0,1}=W^{(1,1)}=6$, $%
W_{1,1}=9.805784090$, $W_{2,1}=13.66928892$ for the three lowest
eigenvalues. In this case the truncation condition only provides the lowest
eigenvalue and misses all the other ones.

According to the author's equations (27) and (28) the energies $\mathcal{E}%
_{1,l}$ depend on $\left| \tilde{Q}_{0}E_{0}\right| $; however, this
symmetry only takes place for the particular states stemming from the
truncation condition. To illustrate this point we have calculated the three
lowest eigenvalues for $\alpha =\alpha _{1,0}^{(2)}=\sqrt{2}$ and $\alpha
_{1,1}^{(2)}=\sqrt{6}$; in the former case we have $W_{0,0}=-1.459587134$, $%
W_{1,0}=W^{(1,0)}=4$, $W_{2,0}=8.344349426$ and $W_{0,1}=1.600357154$, $%
W_{1,1}=W^{(1,1)}=6$, $W_{2,1}=10.21072810$ in the latter. We appreciate
that all the eigenvalues depend on the sign of $\alpha $ except the
particular one given by the truncation condition that in these two cases is
the second lowest.

\begin{figure}[tbp]
\begin{center}
\includegraphics[width=9cm]{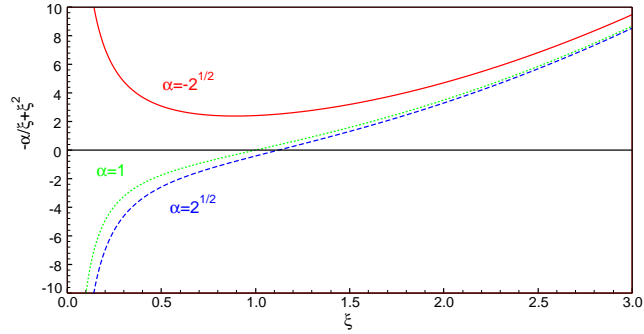}
\end{center}
\caption{Potential-energy functions for three of the $\alpha$ values
discussed in the text}
\label{fig:potential}
\end{figure}

In the examples analized above we have chosen values of $\alpha $ that stem
from the truncation condition. In order to show that the eigenvalue equation
(\ref{eq:dif_eq_R_2}) supports bound states for any value of $\alpha $ we
choose $\alpha =1$ and $l=0$ that are not consistent with the truncation
condition. A straightforward calculation with the methods just mentioned
yields: $W_{0,0}=-0.2085695649$, $W_{1,0}=4.601041510$, $W_{2,0}=8.834509671$
for the three lowest eigenvalues. We want to be clear about this point: for
a given quantum-mechanical model with $\alpha =\alpha _{n,l}^{(i)}$ the
truncation condition only yields one eigenvalue $W^{(n,l)}$ and misses the
rest of the spectrum. For a quantum-mechanical model with $\alpha \neq
\alpha _{n,l}^{(i)}$ that condition does not give any information about its
bound states. This fact is well known by anybody familiar with conditionally
solvable quantum-mechanical eigenvalue equations\cite{D88, BCD17,CDW00}.
Figure~\ref{fig:potential} shows the potential-energy function $-\alpha /\xi
+\xi ^{2}$ for $\alpha =-\sqrt{2},1,\sqrt{2}$, where we appreciate that
there is nothing that forbids the existence of bound states for $\alpha =1$
and the same conclusion can be drawn for any other value of $-\infty <\alpha
<\infty $. It is clear that there is no quantization of $\alpha $ and,
consequently, no quantization of the oscillator frequency $\omega $. Such
conjecture is an artifact of the truncation condition (\ref
{eq:truncation_cond}).

It follows from the Hellmann-Feynman theorem\cite{CDL77,P68} (and references
therein) that the actual eigenvalues $W$ of the differential equation (\ref
{eq:dif_eq_R_2}) decrease with $\alpha $ according to
\begin{equation}
\frac{\partial W}{\partial \alpha }=-\left\langle \frac{1}{\xi }%
\right\rangle ,
\end{equation}
so that one expects negative eigenvalues for sufficiently large values of $%
\alpha $. This conclusion is verified by the calculation discussed above.
Notice that the truncation condition does not predict negative eigenvalues
which clearly shows that one cannot obtain the spectrum of any such model
from the two equations (\ref{eq:truncation_cond}).

Figure~\ref{fig:Wn0} shows several eigenvalues $W^{(n,0)}$ (red circles)
joined by blue lines that mark the curves $W_{\nu ,0}(\alpha )$. For a given
value of $\alpha $ the spectrum of the model is given by the intersection of
a vertical line and the blue lines (for example, the green, dashed line at $%
\alpha =-\sqrt{2}$). Any such vertical line will not meet more than one red
circle, except the one at $\alpha =0$. This fact shows that the truncation
condition only yields the spectrum of the harmonic oscillator ($\alpha =0$).
This figure also shows present results for $\alpha =1$ (blue squares).

Summarizing: Bakke fails to mention that none of the models discussed in his
paper supports bound states because the motion of the particle along the $z$
axis is unbounded. Bound states occur only if we restrict the motion to the $%
x-y$ plane. In the second model (under the latter assumption) the approach
followed by the author only yields some extremely particular states for also
extremely particular values of the model parameter $\alpha $. Anybody
familiar with the Schr\"{o}dinger equation realizes that for any real value
of $\alpha $ in the equation (\ref{eq:dif_eq_R_2}) one obtains an infinite
set of eigenvalues $W_{n,l}(\alpha )$. The truncation method proposed by the
author is unsuitable for obtaining the spectrum except for some particular
values of $\alpha $ and in each of these cases it provides just one
eigenvalue, for example, $W_{0,0}\left( \pm \sqrt{2}\right) =4$ and $%
W_{1,1}\left( \pm \sqrt{6}\right) =6$, and certainly misses all the other
eigenvalues for those particular values of $\alpha $. Present results also
show that there are bound states for values of $\alpha \neq \alpha
_{n,l}^{(i)}$ and for this reason the parameter $\alpha $ or the oscillator
frequency $\omega $ are not quantized as conjectured by the author.
Consequently, the predicted existence of allowed cyclotron frequencies is
merely an artifact of the truncation method proposed by Bakke. Such
conclusion comes from a misinterpretation of the meaning of the exact
solutions to conditionally solvable quantum-mechanically problems\cite
{D88,CDW00,BCD17}. Such solutions are not the only bound states of the
system but just some rather rare eigenfunctions with polynomial factors.

\section*{Addendum}

In this addendum we address the points raised by Bakke\cite{B20} in his
Reply to present Comment. We will restrict ourselves to the most relevant
ones.

\begin{figure}[tbp]
\begin{center}
\includegraphics[width=9cm]{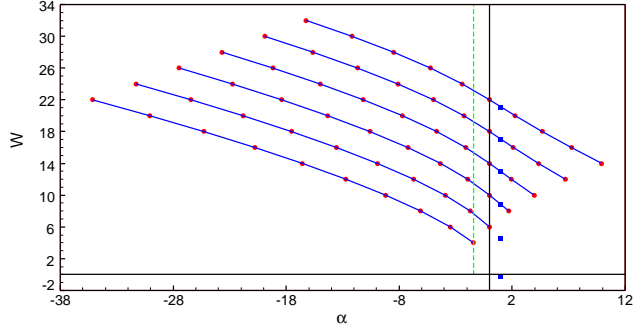}
\end{center}
\caption{Eigenvalues $W_{n,0}$ obtained from the truncation condition (red
circles) and by means of the variational method (blue squares)}
\label{fig:Wn0}
\end{figure}

To begin with, we analyze the following sentences: ``With respect to the
comment about the possible values on the angular frequency $\omega $, what
we have in Ref. 1 is a choice of a parameter that can be adjusted in order
to obtain a polynomial solution to the biconfluent Heun equation. Other
choices could be made, for instance, the parameter $E_{0}$ (Note that $Q$
cannot be used.) However, by analyzing Eq. (26) of Ref. 1, we can observe
that $\mathcal{E}_{n,l}$ is not explicitly dependent on the parameter $E_{0}$%
. This makes to be difficult to solve the coupled equations that stems from
the conditions $g=2n$ and $a_{n+1}=0$ given in Eq. (25) of Ref. 1. The
choice of $E_{0}$ may not exist in the domain of valid solutions to the
biconfluent Heun equation.'' It seems that Bake does not understand the
problem. We have shown that the actual eigenvalues can be written as
\begin{equation}
\mathcal{E}_{\nu ,l}=\frac{\omega }{2}W_{\nu ,l}+\frac{k^{2}}{2m}.
\end{equation}
Since $W_{\nu ,l}$ is a continuous function of $\alpha $, as shown in figure~%
\ref{fig:Wn0} for $l=0$, and $\alpha $ depends on $Q$, $E_{0}$ and $\omega $%
, there is no difference, in principle, in choosing one parameter or the
other. But the most important fact is that since there are bound states in
the $x-y$ plane for all values of $\alpha $ then there are bound states in
the plane \textit{for all values of} $\omega $.

Bakke also states that ``It is worth citing that the same aspect was
observed by Ver\c{c}in$^{19}$ where the exact solutions of the problem of
two anyons subject to a uniform magnetic field in the presence of the
Coulomb repulsion are obtained for discrete values of the magnetic field.
Thereby, the use of the Frobenius method in Ref. 1 is correct and agrees
with Refs. 2, 5, 14, 19-21.'' We agree with Bakke\cite{B20} that Ver\c{c}in%
\cite{V91} conjectured the existence of allowed magnetic-field intensities.
However, Bakke appears to be unaware of the fact that Myrheim et al\cite
{MHV92} wrote a sequel in which they showed that the eigenvalues are in fact
continuous functions of the magnetic-field intensity (see their figure 2 and
the definition of the parameter $b$). In this second paper (coauthored with
Ver\c{c}in) they no longer spoke of allowed values of the field intensity
and even showed how to obtain \textit{all the eigenvalues numerically} by
means of the three-term recurrence relation. They also argued that the
polynomial solutions are just particular cases. The results of Myrheim et al%
\cite{MHV92} are similar to ours (if $4\nu =W$) , except that they have a
Coulomb term of the form $b/\rho $ that leads to a positive slope for $\nu
(b)$.

Let us now analyze the following sentences: ``The main discussion proposed
by Fern\'{a}ndez is the use of Rayleigh-Ritz variational method in search of
approximate solutions to the Schr\"{o}dinger equation. The trial wave
function used by Fern\'{a}ndez does not corresponds to the wave function
given in terms of the polynomial solution to the biconfluent Heun equation
obtained in Ref. 1. The analysis made in Ref. 1 is based on constructing a
polynomial of first degree for the biconfluent Heun function. From Eq. (22)
of Ref. 1, we obtain: $H(\xi )=a_{0}+a_{1}\xi $. Note $a_{0}=1$ and $%
a_{1}\neq 0$ as shown in Eqs. (24) of Ref. 1. Since the trial function used
by Fern\'{a}ndez is not given in terms of this polynomial solution, the
comparison between the methods is not clear. Furthermore, as we can see in
Ver\c{c}in's paper,$^{19}$ the exact solutions exist only for
well-determined values of the magnetic field. By contrast, Fern\'{a}ndez's
comment shows determined values of the parameter $W$ or the parameter $E_{0}$%
. This makes to be difficult to compare the result obtained by Fern\'{a}ndez
because the choice of $E_{0}$ may not exist in the domain of valid solutions
to the biconfluent Heun equation. In this way, a simple comparison, as
Fern\'{a}ndez tries to make, is very simple and has no mathematical and
physical basis for such a comparison.''

It seems that Bakke is not familiar with the Ritz variational method. We
have carried out the calculation with $N\leq 15$ Gaussian functions and
verified that they converged from above to the last digit reported in the
Comment. Since the basis set of Gaussian functions is complete we are
confident that the results are correct, as anybody can verify because the
calculation is extremely simple. When $\alpha $ takes one of the values
given by the truncation condition the Ritz variational method yields the
value of $W$ given by that condition and the expansion reduces to the
polynomial function mentioned by Bakke. We think that this fact is clearly
shown in the Comment, as well as that the Ritz method also yields other
eigenvalues omitted by the truncation condition. However, in order to
address Bakke's criticisms, in this Addendum we add tables \ref{tab:alpha1}
and \ref{tab:alpha2} that show the convergence of the Ritz variational
method for $l=0$, $\alpha =-\sqrt{2}$ and $\alpha =\sqrt{2}$. It is worth
noticing that the roots of the secular determinant approach to the
eigenvalues from above, except in the case of the root $W=4$ of the
truncation condition where the Ritz variational method yields the exact
eigenvalue from $N=2$ on. The reason is that in this case the Ritz trial
function reduces to a polynomial of degree one as argued by Bakke\cite{B20}.
More precisely, the Ritz variational method yields the eigenvalues coming
from the Frobenius method as well as all those eigenvalues that the
truncation condition fails to provide. We have already referred to Ver\c{c}%
in's conclusion above that was proved wrong in a sequel. What we fail to
understand is why Bakke states that we analyzed the behaviour of $W$ with
respect to $E_{0}$. Our figure 2 clearly shows that we calculated $W$ as
function of $\alpha $, which depends on $E_{0}$ but also on $Q$ and $\omega $%
. This parameter was defined by Bakke and we used it exactly as it appears
in the Coulomb term in his paper\cite{B14b}.

\begin{table}[tbp]
\caption{Lowest variational eigenvalues $W_{\nu,0}$ for $l=0$ and $\alpha=-%
\protect\sqrt{2}$}
\label{tab:alpha1}
\begin{center}
\par
\begin{tabular}{D{.}{.}{3}D{.}{.}{11}D{.}{.}{11}D{.}{.}{11}D{.}{.}{11}}
\hline \multicolumn{1}{c}{$N$}& \multicolumn{1}{c}{$W_{0,0}$} &
\multicolumn{1}{c}{$W_{1,0}$} & \multicolumn{1}{c}{$W_{2,0}$}  & \multicolumn{1}{c}{$W_{3,0}$}\\
\hline

 2  &  4.000000000  &   10.49997602  &                &                \\
 3  &  4.000000000  &   7.751061995  &   19.88102859  &                \\
 4  &  4.000000000  &   7.694010921  &   11.97562584  &   33.92039998  \\
 5  &  4.000000000  &   7.693979367  &   11.51212379  &   17.05520450  \\
 6  &  4.000000000  &   7.693978905  &   11.50604696  &   15.46896992  \\
 7  &  4.000000000  &   7.693978892  &   11.50604243  &   15.37652840  \\
 8  &  4.000000000  &   7.693978891  &   11.50604238  &   15.37592761  \\
 9  &  4.000000000  &   7.693978891  &   11.50604238  &   15.37592718  \\
10  &  4.000000000  &   7.693978891  &   11.50604238  &   15.37592718  \\
\end{tabular}
\par
\end{center}
\end{table}

\begin{table}[tbp]
\caption{Lowest variational eigenvalues $W_{\nu,0}$ for $l=0$ and $\alpha=%
\protect\sqrt{2}$}
\label{tab:alpha2}
\begin{center}
\par
\begin{tabular}{D{.}{.}{3}D{.}{.}{11}D{.}{.}{11}D{.}{.}{11}D{.}{.}{11}}
\hline \multicolumn{1}{c}{$N$}& \multicolumn{1}{c}{$W_{0,0}$} &
\multicolumn{1}{c}{$W_{1,0}$} & \multicolumn{1}{c}{$W_{2,0}$}  & \multicolumn{1}{c}{$W_{3,0}$}\\
\hline

 2&  -1.180391283 &  4.000000000 &              &               \\
 3&  -1.401182256 &  4.000000000 &  9.284143096 &               \\
 4&  -1.449885589 &  4.000000000 &  8.345259771 &  17.66452696  \\
 5&  -1.458156835 &  4.000000000 &  8.344361267 &  12.69095166  \\
 6&  -1.459389344 &  4.000000000 &  8.344349784 &  12.53313315  \\
 7&  -1.459560848 &  4.000000000 &  8.344349442 &  12.53290257  \\
 8&  -1.459583736 &  4.000000000 &  8.344349427 &  12.53290132  \\
 9&  -1.459586704 &  4.000000000 &  8.344349427 &  12.53290130  \\
10&  -1.459587081 &  4.000000000 &  8.344349427 &  12.53290130  \\
11 &  -1.459587128 & 4.000000000 & 8.344349427 & 12.53290130  \\
12 &  -1.459587134 & 4.000000000 & 8.344349427 & 12.53290130  \\
13 &  -1.459587134 & 4.000000000 & 8.344349427 & 12.53290130  \\
\end{tabular}
\par
\end{center}
\end{table}


\begin{thebibliography}{99}
\bibitem{B14b}  K. Bakke, Int. J. Mod. Phys. A \textbf{29}, 1450117 (2014).

\bibitem{F20}  F. M. Fern\'{a}ndez, Dimensionless equations in
non-relativistic quantum mechanics, arXiv:2005.05377 [quant-ph].

\bibitem{LL65}  L. D. Landau and E. M. Lifshitz, Quantum Mechanics.
Non-relativistic Theory (Pergamon, New York, 1958).

\bibitem{CDL77}  C. Cohen-Tannoudji, B. Diu, and F. Lalo\"{e}, Quantum
Mechanics (John Wiley \& Sons, New York, 1977).

\bibitem{D88}  A. DeSousa Dutra, Phys. Lett. A \textbf{131}, 319 (1988).

\bibitem{BCD17}  S. Bera, B. Chakrabarti, and T. K. Das, Phys. Lett. A
\textbf{381}, 1356 (2017).

\bibitem{CDW00}  M. S. Child, S-H. Dong, and X-G. Wang, J. Phys. A \textbf{33%
}, 5653 (2000).

\bibitem{P68}  F. L. Pilar, Elementary Quantum Chemistry (McGraw-Hill, New
York, 1968).

\bibitem{FMT89a}  F. M. Fern\'{a}ndez, Q. Ma, and R. H. Tipping, Phys. Rev.
A \textbf{39}, 1605 (1989).

\bibitem{B20}  K. Bakke, Int. J. Mod. Phys. A \textbf{25}, 2075003 (2020).

\bibitem{V91}  A. Ver\c{c}in, Phys. Lett. B \textbf{260}, 120 (1991).

\bibitem{MHV92}  J. Myrheim, E. Halvorsen, and A. Ver\c{c}in, Phys. Lett. B
\textbf{278}, 171 (1992).
\end{thebibliography}
\end{document}